# The Affinity of the Sulfate- and Ether-Containing Surface-Active Ionic Liquids to Carbon Dioxide, Hydrogen Fluoride, Hydrogen Sulfide, and Water

Vitaly V. Chaban[1,2]

(1) Federal University of Sao Paulo, Sao Paulo, Brazil.

(2) P.E.S., Vasilievsky Island, Saint Petersburg 190000, Russian Federation.

**Abstract.** The development of novel task-specific ionic liquids (ILs) represents an essential challenge in modern organic and physical chemistries. Recently we reported surface-active ILs contained the two well-known organic cations (1-butyl-3-methylimidazolium [bmim$^+$] and tetrabutylammonium [TBA$^+$]) and the two surface-active anions (lauryl sulfate [C$_{12}$SO$_4^-$], lauryl ether sulfate [C$_{12}$ESO$_4^-$]). In the present work, we investigate the affinity of these ionic compounds to the selected small molecules that exhibit practical implications: water, hydrogen fluoride, hydrogen sulfide, and carbon dioxide. We identified that the sulfate group, the ether groups, and the aromatic imidazole ring make the strongest contributions to the physical sorption of the polar gas molecules. In turn, the [TBA$^+$] cation, the saturated hydrocarbon chain of the anions, and the alkyl chains of [bmim$^+$] contribute to a significantly smaller extent. The reported data are interesting in the context of using surface-active ILs in the oil industry to capture and store undesirable and toxic gases.

**Introduction**

Surface-active ionic liquids (SAILs) constitute an interesting class of ionic compounds with peculiar physicochemical properties.[1-13] During the last years, quite a few SAILs have been introduced and their potential applications have been considered. Some SAILs contain long hydrophobic (tensoactive) chains in the structure of their cations, anions, or both of them.[14-16] In the meantime, most cations used to synthesize an ionic liquid possess amphiphilic moieties that modulate solvation behavior and phase transition points.[17] The cations whose alkyl chains are longer than six carbon atoms act as emulsifiers. Therefore, surface activity is acquired by the resulting ionic liquids. A knowledgeable combination of hydrophobic and hydrophilic structures of SAILs allows improving their solubility in hydrophobic environments.[16]

The known SAILs are thermodynamically stable compounds that tend to self-organize in solutions giving rise to liquid crystals, micelles, and vesicles.[14-15, 18-20] At water contents exceeding a certain proportion, the formation of gels was reported and rationalized.[19] Experimental efforts must be devoted to investigating the physicochemical properties of SAILs in different media as a function of their concentrations and temperature.

As of today, ionic liquids are actively probed to bind certain gases, such as carbon dioxide.[16, 21-25] The reason is their low volatility, non-flammability, good chemical and thermal stabilities, and sometimes low toxicity to human beings.[26-28] A variety of methods and synthetic modifications were proposed to achieve better sorption performance of ILs. Both physical and chemical sorptions are relevant for specifically tuned structures of ILs.[21, 24, 29-30]

In our recent work, we reported a series of SAILs based on the long-alkyl-chained anions that exhibited interesting properties,[16] such as 1-butyl-3-methylimidazolium lauryl sulfate, 1-butyl-3-methylimidazolium lauryl ether sulfate, tetrabutylammonium lauryl sulfate, and tetrabutylammonium lauryl sulfate. Presently, we document their binding energies to a few important small molecules, such as water, carbon dioxide, hydrogen sulfide, and hydrogen fluoride. The obtained results help understand the prospects of employing these SAILs and their aqueous solutions for energy-efficient physisorption.

**Computational Methodology**

Binding energies, dipole moments, and geometric parameters were computed for the relaxed geometries. The geometry convergence criterion upon relaxation was set to $10^{-4}$ Hartree (BFGS algorithm). If the energy alteration between the two subsequent geometries was smaller than the designated threshold, the local minimum was considered to be located. Vibrational frequency analysis was used to exclude transition states. The wave function was constructed in accordance with hybrid density functional theory, Becke-3-Lee-Yang-Parr (B3LYP).[31-32] This exchange-correlation functional is one of the most accurate and successful generalized-gradient solutions to simulate organic and inorganic, neutral, and ionic species. The atom-centered split-valence triple-zeta basis set and polarization and diffuse functions were employed, 6-311+G*. The basis set superposition error was deducted from the reported binding energies by using the counterpoise method composed of the three consequent calculations. GAMESS-US was used to perform single-point calculations[33] and SciPy (scipy.org) functions were used to construct and navigate[34-36] the potential energy surface of the investigated complexes.[37] Gabedit 2.8 was used to construct initial molecular geometries.[38]

**Results and Discussion**

Binding energies (Table 1) determine the strength of the molecule-molecule or molecule-ion interactions. Decomposition of the total binding energy into components allows understanding, which moiety of ILs contributes most significantly to the gas capturing performance. Apart from $CO_2$, we also computed binding energies for $H_2O$, $H_2S$, and HF (Table 1). Fixation of $H_2S$ and HF is relevant in the oil industry, whereas interaction with $H_2O$ determines the aqueous solubility of these ILs. More polar molecules generally exhibit higher binding energies, while $CO_2$ is most difficult to capture due to its peculiar structure and relative chemical inertness. Table 1 also summarizes dipole moments of the studied complexes for comparison. The results for the ammonium cation are provided to see its difference from $N(C_4H_9)_4^+$. Although $NH_4^+$ exhibits much better binding energies to the water and gas molecules, it does not give rise to any IL due to strong cation-anion coordination.

The ether group exhibits a twice weaker affinity to $CO_2$ (10 kJ mol$^{-1}$), as compared to the sulfate group (21 kJ mol$^{-1}$), but its overall impact is significant since the $C_{12}ESO_4^-$ anion synthesized by our team previously[16] has approximately six ether groups. Furthermore, the ether groups are not sterically blocked, unlike $C_{12}ESO_4^-$, due to the cation-anion coordination in the concentrated solutions of [bmim][$C_{12}ESO_4$] and [TBA][$C_{12}ESO_4$]. The bmim$^+$ cation is more efficient than TBA$^+$, 14 vs. 7 kJ mol$^{-1}$, but the advantage of TBA$^+$ is the absence of available electron-deficient sites. Therefore, TBA$^+$-containing ILs are expected to be generally more efficient assuming the same anion. HF can slightly polarize the hydrophobic chain providing 4 kJ mol$^{-1}$, while interaction is virtually absent in all other cases. Since the hydrocarbon chain is useless for gas capture, its role is only regulation of the phase transition points of the corresponding ILs. Phase transitions can also be tuned by a number of the -CH$_2$-O- fragments, therefore maintenance of a long hydrocarbon chain during further engineering does not look critical.

Table 1. The selected characteristics of the complexes of the selected moieties of the ILs and small molecules (carbon dioxide, water, hydrogen fluoride, hydrogen sulfide): binding energies, dipole moments, minimum atom-atom distances, and non-covalent angles.

| Neutral or charged moiety | Small molecule | Binding energy, -$E_b$, kJ mol$^{-1}$ | Dipole moment, D | Minimum distance, Å | Angle*, degree |
| --- | --- | --- | --- | --- | --- |
| CH₃SO₄⁻ | CO₂ | 21 | 5.2 | 2.66 (O-C) | 93 (O-C-O) |
| | H₂O | 60 | 4.8 | 2.09 (O-H) | 145 (O-H-O) |
| | HF | 91 | 4.3 | 1.58 (O-H) | 173 (O-H-F) |
| | H₂S | 40 | 4.1 | 1.91 (O-H) | 171 (O-H-S) |
| ether | CO₂ | 10 | 1.9 | 2.75 (C-O) | 92 (O-C-O) |
| | H₂O | 23 | 3.5 | 1.88 (O-H) | 180 (O-H-O) |
| | HF | 45 | 4.6 | 1.65 (O-H) | 177 (O-H-F) |
| | H₂S | 11 | 3.3 | 2.09 (O-H) | 180 (O-H-S) |
| saturated hydrocarbon chain | CO₂ | 0 | 0.0 | 4.25 (O-H) | 93 (H-O-C) |
| | H₂O | 1 | 2.6 | 2.46 (O-H) | 125 (H-O-H) |
| | HF | 4 | 2.8 | 2.19 (H-H) | 158 (H-H-F) |
| | H₂S | 0 | 1.5 | 2.79 (H-H) | 159 (H-H-S) |
| bmim⁺ | CO₂ | 14 | 5.4 | 2.31 (O-H) | 163 (H-O-C) |
| | H₂O | 45 | 5.2 | 2.05 (O-H) | 125 (H-O-H) |
| | HF | 31 | 4.9 | 2.15 (F-H) | 155 (H-F-H) |
| | H₂S | 19 | 6.0 | 2.73 (S-H) | 114 (H-S-H) |
| NH₄⁺ | CO₂ | 39 | 11 | 1.84 (O-H) | 180 (H-O-C) |
| | H₂O | 96 | 2.7 | 1.65 (O-H) | 126 (H-O-H) |

| | HF | 59 | 3.6 | 1.75 (F-H) | 180 (H-F-H) |
| --- | --- | --- | --- | --- | --- |
| | H$_2$S | 57 | 7.2 | 2.22 (S-H) | 103 (H-S-H) |
| N(C$_4$H$_9$)$_4$$^+$ | CO$_2$ | 7 | 2.8 | 2.67 (O-H) | 157 (H-O-C) |
| | H$_2$O | 27 | 1.9 | 2.32 (O-H) | 118 (H-O-H) |
| | HF | 19 | 1.6 | 2.36 (F-H) | 180 (H-F-H) |
| | H$_2$S | 10 | 1.2 | 3.20 (S-H) | 133 (H-S-H) |

\* The angle is formed by an atom of the investigated moiety, which performs coordination, a coordinated atom of the small molecule, and an atom covalently bound to the latter.

**Conclusions**

In the present work, we report a large set of binding energies to characterize interactions of novel surface-active ILs with selected polar gases and water. We show that the polar moieties of 1-butyl-3-methylimidazolium lauryl sulfate, 1-butyl-3-methylimidazolium lauryl ether sulfate, tetrabutylammonium lauryl sulfate, and tetrabutylammonium lauryl sulfate exhibit significant binding affinity to hydrogen fluoride, hydrogen sulfide, carbon dioxide, and water. The investigated surface-active ILs may represent interest to the oil industry as compounds that can physically adsorb undesirable small molecules.

The present work confirms our previously published hypothesis[16] that binding of [bmim$^+$] to water is stronger as compared to binding of [TBA$^+$] to water in all researched surface-active ILs thanks to an acidic hydrogen atom of the imidazole ring. The emerged H-bonding is responsible for gel and liquid crystal formation. The [C$_{12}$SO$_4$$^-$] and [C$_{12}$ESO$_4$$^-$] anions differ by only the presence or absence of a few ether groups in their chains. Hereby we rationalize and corroborate the improvement of the aqueous solubility due to the presence of the ether groups. The presently computed binding energies explain why the [TBA][C$_{12}$SO$_4$] IL is immiscible with water as revealed experimentally.[16]

The investigated SAILs are prospective inexpensive scavengers of CO$_2$, H$_2$S, and HF. Therefore, some of them potentially deserve practical implementation in the oil industry in the form of aqueous solutions.